# Geometrical lattice engineering of complex oxide heterostructures: a designer approach to emergent quantum states

Xiaoran Liu, S. Middey, Yanwei Cao, M. Kareev, and J. Chakhalian*

Department of Physics, University of Arkansas, Fayetteville, Arkansas 72701, USA

*Address all correspondence to J. Chakhalian at jchakhal@uark.edu

## Abstract

Epitaxial heterostructures composed of complex oxides have fascinated researchers for over a decade as they offer multiple degrees of freedom to unveil emergent many-body phenomena often unattainable in bulk. Recently, apart from stabilizing such artificial structures along the conventional [001] direction, tuning the growth direction along unconventional crystallographic axes has been highlighted as a promising route to realize novel quantum many-body phases. Here we illustrate this rapidly developing field of geometrical lattice engineering (GLE) with the emphasis on a few prototypical examples of the recent experimental efforts to design complex oxide heterostructures along the (111) orientation for quantum phase discovery and potential applications.

## I. Introduction

The search and exploration of new collective quantum states is of prime importance and interest in condensed matter physics. Towards this goal, ultra-thin heterostructures composed of two or more structurally, chemically and electronically dissimilar constituent oxides have been developed into a powerful approach over the past decade. [1-6] The main notion here is that at the interface where the dissimilarities meet, the frustration caused by mismatches between arrangement of atoms, charges, orbitals or spins can trigger the emergence of phenomena with electronic and magnetic structures markedly different from the corresponding bulk compositions. [1] As a result, the interface engineering (IE) has opened a route to novel material behaviors by means of those mismatches as the control parameters. The IE approach is intimately connected to another popular approach to tailor the properties of materials with epitaxial strain by effectively altering the bond-length and bond-angle of structural units through the deliberate choice of substrates. The exploration of epitaxial strain due to the lattice mismatch has been thus far successfully used to manipulate the electronic bandwidth, band filling, ferroelectric and magnetic interactions of the ultra-thin films. [7-9]

Inspired by the success of those engineering methods, very recently another promising venue collectively known as geometrical lattice engineering (GLE) has been presented as a powerful tool to forge new topological and quantum many-body states. [10-12] In close synergy with the interface engineering (IE) and strain engineering (SE) where mismatches between layers can induce unusual interactions, the key idea behind the GLE is to design fully epitaxial ultra-thin heterostructures with an artificial lattice geometry generated by stacking of a very *specific number of atomic planes along a* *specific orientation*. This concept can be further illustrated by realizing that for a three-dimensional (3D) material the stacking of two-dimensional (2D) atomic planes, the specific arrangement of ions in those planes, their sequence, and the periodicity of layers fulfilling a complete unit cell can exhibit drastic variations depending on the crystallographic direction along which it is projected. Conventionally, for rather thick bulk like films, the effect of those variations is often negligible (apart from anisotropy), whereas for the ultra-thin samples it becomes much more dominant in determining the electronic and magnetic properties. Following this idea, in the pursuit of exotic quantum states many interesting material systems have been proposed theoretically [10-22], while the experimental work on GLE has been primarily focused on growth of cubic or pseudo-cubic (111)-oriented artificial lattices [23-28].

The rest of the paper is organized as follows. First, we will illustrate the concept of GLE by describing several prototypical examples with (111) orientation to shed light on the details of their design, fabrication, and possible growth challenges. By no means does this paper present a comprehensive review of GLE. We, and as such the authors, have not attempted to review the large body of published results. Instead, we have focused on a few selected results to illustrate the concepts, methodologies and physics behind described phenomena. We apologize for possible omissions. In the final section, a brief outlook will be provided to accentuate some intriguing new ideas and material systems with other lattice orientations and highlight the significant but largely unexplored potential of GLE for quantum phase discovery and applications.

## II. Control parameters of GLE

In general, the GLE concept is comprised of three controllable stages throughout the process of heteroepitaxial fabrication. Here, we use a 3D simple cubic unit cell model to schematically illustrate these control parameters, as shown in Fig. 1.

A. *Growth orientation*: Starting with the same bulk compound, its 3D crystal structure is viewed as a stacking of atomic layers with different in-plane lattice geometries along different directions. For example, as shown in Fig. 1, while the (001) planes have square symmetry, the (110) and (111) planes have rectangular and triangular geometry, respectively. The required in-plane lattice geometry by design can be thus determined by selecting the proper structure and orientation of the substrate surface, which acts as a guiding template during the initial stage of nucleation and growth. A typical example for this case is the realization of 2D magnetically frustrated lattices derived from ultra-thin (111)-oriented spinel-type structure $AB_2O_4$ (for more details see section III).

B. *Out-of-plane stacking sequence*: In a bulk crystal the periodicity of the atomic layers to fulfill the requirement of translational symmetry as well as the relative atomic positions in neighboring lattice planes vary dramatically with the specific choice of the crystallographic direction. For instance, the stacking of the adjacent layers can be either right on top of each other [the (001) stacking in Fig. 1], or shifted [the (110) stacking], or even entirely inversed [the (111) stacking]. This observation is of paramount importance for the design of artificial heterostructures since by controlling the number of stacking layers within that period one can devise unique quasi-2D lattices that cannot be obtained in the naturally formed crystals. Among the prominent example of GLE we cite the generalized graphene lattice, which can be obtained by digitally tuning the number of atomic layers of (111)-oriented $ABO_3$ perovskite-type structures for searching potential topological phases with and *without* spin-orbit active ions (for more details see section IV).

C. *Isostructural superlattices approach:* Combining isostructural materials together to establish superlattice structures via digital control over the individual number of layers adds another practical dimension to the application of GLE. This approach can be very useful for the purpose of achieving materials with complex chemical compositions or even thermodynamically unstable phases in the bulk form. A representative test case is the fabrication of (111)-oriented $1ABO_3/1AB'O_3$ superlattices (here '1' refers to single pseudocubic/cubic unit cell), giving rise to $A_2BB'O_6$ double-perovskite. Further examples can be found in section V.

In what follows we discuss in some details how the control parameters of GLE can be experimentally realized in the fabrication of complex oxide thin films and superlattices.

## III. Geometrically frustrated lattices

Various geometrically frustrated magnetic systems including spin ice, spin glass and spin liquid have been studied over several decades. [29-34] In such systems, the exchange interaction between nearest neighbor ions is typically antiferromagnetic (for spin ice, the exchange interaction is ferromagnetic) and the peculiar geometries of the lattices (triangular and kagome lattice for 2D case, or pyrochlore and hyper-kagome lattice for 3D case) render the magnetic bonds impossible to be satisfied simultaneously. These geometrically frustrated lattices, especially the 2D ones have an excellent potential to host various intriguing exotic phenomena such as quantum spin liquid, topological phases, kinetic ferromagnetism, and chiral superconducting state to list a few. [16,35-37] The fabrication of a 2D geometrically frustrated lattice, however, has proven to be rather challenging. For example, even though several metal-organic hybrid counterparts were synthesized by chemical methods, structural disorder and strong lattice distortions remained a persistent hindrance. [38,39] Alternatively, the GLE approach offers a promising way to generate such frustrated lattice by growing ultra-thin spinels along the [111] crystallographic axis, as shown in Fig. 2.

First we recap that the general chemical formula of the spinel structure can be written as $A_{1-\delta}B_{\delta}[A_{\delta}B_{2-\delta}]O_4$; with $\delta = 0$, it is known as normal spinel $AB_2O_4$, in which all A-site cations are tetrahedrally coordinated while all B-site cations are octahedrally coordinated. Next we point out that the normal spinel shows a structurally unique feature when viewed along the [111] crystallographic axis. Specifically, the crystal structure can be regarded as a stacking of alternative lattice planes each of which is composed of single kind of ions (see Fig. 2). In this stacking the O sublattice has the cubic close packing (CCP) atomic arrangement. In contrast, the A and B ionic planes show three distinct types of geometrically frustrated lattices, namely, triangular A plane (T plane), triangular B plane (T' plane), and kagome B plane (K plane) as displayed in Fig. 2. In the bulk, totally 18 ionic planes complete a full periodicity in the form of $[\text{O-K-O-T-T'-T}]_3$ [40] with [O-K-O-T-T'-T] as the basic repeat unit. For the rest of this paper such a basic repeat unit is referred as "1 monolayer" of the (111)-oriented normal spinel.

From the synthesis point of view, there are two main issues one needs to overcome to obtain high quality (111)-oriented normal spinel heterostructures. First, due to the limited availability of isostructural/isosymmetric substrates, large lattice mismatch (> 5%) between the film and substrate generally leads to incoherent or inhomogeneous growth. Secondly, a magnetically passive spacer in the superlattice has to be chemically stable and lattice symmetry compatible with the spinel structure. A complete solution to these issues has been recently demonstrated by Liu *et al.* [24] in the epitaxial growth of (111)-oriented $CoCr_2O_4/Al_2O_3$ (CCO/AlO) heterostructures. For this system, AlO was selected as both the substrate and spacer because (1) AlO is a non-magnetic insulator with a large band-gap of about 9 eV and (2) since along the (0001) direction the oxygen sublattice consists of hexagonal close packing (HCP) arrangement, it has the in-plane lattice geometry which almost perfectly matches the CCP arrangement of spinel (see Fig. 3(a)). In addition, a fairly small lattice mismatch of ~ 3% can be obtained when we compare the respective in-plane O-O distances (AlO: 2.74 Å; CCO: 2.82 Å).

The spinel-based heterostructures were fabricated layer-by-layer using the pulsed laser deposition (PLD) technique. The structural properties of the resultant samples were characterized by X-ray reflectivity (XRR) and diffraction (XRD) measurements with the results for two prototypical samples (a thicker $[2CCO/2AlO]_{10}$ and an ultra-thin $[2CCO/1AlO]_4$) shown in Fig. 4. As seen, according to the XRR scans with fittings from Fig. 4(a), 1CCO has a thickness of ~ 4.93 Å while 1AlO ~ 13.5 Å and the average interface roughness of about 3 Å. On the other hand, the XRD 2θ-ω scans shown in Fig. 4(b) in the vicinity of the substrate (006) reflection confirm that the films are grown along the (111) direction with no additional peaks from impurity phase. In particular, besides the film's (222) main reflection (labeled as $0^{th}$), the superlattice's satellite peaks labeled as $\pm 1^{st}$ were clearly distinguishable and the number of the thickness fringes between the main peak and the satellite peak agrees quite well with the superlattice repeat pattern. These observations provide a solid evidence that growth of the geometrically engineered spinel-type structures along (111) is experimentally feasible.

Another important issue is related to cation disorder or a mixture of various cation valences that can occur during the growth of spinel films [41-46] and if present it may potentially destroy the expected normal spinel stacking patterns. Thus, it was necessary to investigate the chemical states and environments of both Co and Cr ions in the films. Such information can be readily probed by soft X-ray absorption spectroscopy (XAS) measurements. This synchrotron-based technique is extremely sensitive to the local electronic properties of a specific chemical element. The dataset was collected on the $L_{3,2}$ absorption edge of both Co and Cr elements of a representative 2CCO/2AlO superlattice in the total electron yield. As seen in Fig. 4(c) and 4(d), both the Co and Cr XAS of the 2CCO/2AlO sample have lineshapes (including the peak positions as well as the fine multiplet features) almost identical to those of the CCO bulk samples implying the absence of any detectable cation distribution disorder or mixture of chemical valences. This result further confirmed that the normal spinel structure was indeed maintained in the superlattice (i.e. $Co^{2+}$ ions were tetrahedrally coordinated while $Cr^{3+}$ ions were octahedrally coordinated).

It is worthy of noting that in addition to sandwiching the ultrathin spinel structure towards the 2D limit with magnetically inactive spacers [24,47], it may be also interesting to explore (111)-oriented all magnetic spinel superlattices. To illustrate this idea we note that since magnetic properties of spinels strongly depend on the exchange interaction between A and B sites, the periodic modulation of the exchange interaction in artificial superlattices can give rise to novel properties not existing in either the single composition or the solid solution. Murata *et al.* [48] recently reported the growth of such $ZnFe_2O_4/ZnCr_2O_4$ superlattices on $MgAl_2O_4$ (111) substrates. Both of the constituent layers have a normal spinel structure with non-magnetic A sites ($Zn^{2+}$, S = 0) and magnetic B sites ($Fe^{3+}$, S = 5/2; $Cr^{3+}$, S = 3/2). The effect of GLE was investigated by measuring zero field cooling (ZFC) and field cooling (FC) magnetic susceptibilities of the $ZnFe_2O_4$ and the $ZnCr_2O_4$ thin films, the $Zn(CrFe)O_4$ solid solution, and the $mZnFe_2O_4/mZnCr_2O_4$ (m = 1, 2, 4 and 10) superlattices. In the bulk both materials are characterized by large and negative Curie-Weiss temperatures; the $ZnFe_2O_4$ film shows spin-glass behavior below ~17 K while $ZnCr_2O_4$ film shows antiferromagnetic ordering below 13 K. In contrast to this, the $Zn(CrFe)O_4$ solid solution exhibits a distinct spin-glass behavior below 28 K. This higher transition temperature $T_g$ is attributed to the random distributions of the $Fe^{3+}$ and $Cr^{3+}$ ions on the B sites causing spatially inhomogeneous exchange interactions. On the other hand, the spin-glass behavior was also observed for all of the superlattice films with a non-monotonic variation of $T_g$ with the layer thickness m. These experimental results were interpreted as the result of a periodic modulation of the exchange coupling between the B site ions across the interfaces. As the thickness of both layers becomes smaller, the number of the bulk-like $Cr^{3+}$-O-$Cr^{3+}$ as well as the $Fe^{3+}$-O-$Fe^{3+}$magnetic bonds decreases. At the same time, more $ZnFe_2O_4/ZnCr_2O_4$ interfaces appear where artificially induced antiferromagnetic $Cr^{3+}$-O-$Fe^{3+}$ interaction starts to compete with the bulk-like magnetic interactions. As the result of such competition, the formation of a long range magnetic ordering is suppressed and the spin-glass behavior gradually arises. In addition, once the layer thickness of each composition decreases below a certain critical limit (m = 4), the material can be viewed as mainly being made of the interfaces where the induced $Cr^{3+}$-O-$Fe^{3+}$ interaction is dominant and a new magnetic state dramatically different from the bulk appears in the superlattice. The suppression of the spin-glass behavior for $1 < m < 4$ was interpreted originating from some new type of short-range magnetic clusters induced by the $Cr^{3+}$-O-$Fe^{3+}$ interaction, which again does not exist in the constituent bulk compounds. These examples corroborate that apart from chemical doping and epitaxial strain, geometrical lattice engineering of the stacking layers as well as their sequences in the heterostructures is an effective way to alter electronic and magnetic behaviors.

## IV. Generalized graphene lattice from $ABO_3$ perovskites

Perovskite oxides are the host of diverse properties including metal-insulator transition, magnetism, superconductivity, charge and orbital orderings, multiferroicity etc. These infinite layer $ABO_3$ perovskite compounds consist of alternating $AO/BO_2$, $ABO/O_2$, $AO_3/B$ atomic planes along the pseudocubic (pc) [001], [110], and [111] directions, respectively. Thus the precise control during the growth of two or three pseudocubic unit cells of $ABO_3$ along the (111) orientation leads to new lattice geometries with vertically shifted triangular planes of *B* sites and results in buckled honeycomb lattice [10-12] and dice lattice [49] respectively as shown in Fig. 5. The emergence of striking topological phases including anomalous quantum Hall state was originally predicted in a honeycomb lattice by Haldane [50], and recently initiated active searches for artificially stabilized graphene like quasi-2D lattice that can provide an ideal playground for interacting topological phases in complex oxides [10-12,14,19-21,51-59]. Here we describe the case of rare-earth nickelates that vividly illustrates the opportunities for designer topological phases by GLE.

The first member of rare earth nickelates series, $LaNiO_3$ is a paramagnetic metal. The other members (*RE*$NiO_3$ with *RE* = Pr, Nd…….Lu and Y) in the bulk form exhibit metal-insulator transition, E`-type antiferromagnetic ordering, charge ordering and structural transition with a strong dependence of the transition temperature on the size of *RE* ions [60,61]. The possibility of realizing cuprate-like Fermi surface has led to a rapid development and study of (001) oriented ultra-thin films and superlattices of various members of the *RE*$NiO_3$ family (see [62] for review). Several theoretical works [11,12,51-53] further emphasized the possibility of realizing *interaction*-driven topological phases without spin-order (*e.g.* Dirac half semimetal phase, quantum anomalous Hall insulator phase, or ferromagnetic nematic phase) in the weakly correlated limit on the buckled honeycomb lattice of *RE*$NiO_3$. Moreover, in contrast to the bulk $LaNiO_3$, where orbital ordering is not present, theoretical modeling in strongly correlated limit predicted the presence of an orbitally ordered magnetic phase as the ground state [12,51-53] for the hexagonal $LaNiO_3$.

Despite the conceptual simplicity, growth of perovskites along [111] direction presents formidable challenge. Contrary to the $[001]_{pc}$ direction, the epitaxial stabilization along $[111]_{pc}$ direction is far less explored due to unavoidable surface/chemical reconstruction effects as all perovskite substrates are strongly polar along this direction, e.g. $SrTiO_3$: $[SrO_3]^{4-}$, $Ti^{4+}$ ; $LaAlO_3$:$[LaO_3]^{3-}$, $Al^{3+}$ and so on. While growing a metallic buffer layer at the very beginning of growth may help to avoid this polarity problem by effective screening of dipoles [63], one should pay particular attention as unwanted interfacial (between buffer and desired material) effects may have significant influence on the buffered heterostructure. As the reconstruction effect often appears only at the substrate-vacuum interface, the choice of a substrate with the same charge per plane sequence as that of the desired material can be another solution to this problem, which does not require to grow a buffer layer. To investigate this, Middey *et al*. [64] have grown $LaNiO_3$ films on $SrTiO_3$ (with polar jump at film/substrate interface: see upper panel of Fig. 6(a)) and $LaAlO_3$ (without any polar jump at film/substrate interface: see lower panel of Fig. 6(a)) (111) substrate using pulsed laser interval deposition. The growth was monitored by *in-situ* high pressure RHEED (reflection high energy electron diffraction)**.** It was shown that while thick bulk-like $LaNiO_3$ film is metallic and can effectively screen charge dipoles, a few unit cell thin LNO film is insulating because of quantum confinement [23], and does require additional effects to avoid polar catastrophe. Specifically, in sharp contrast to the film grown on LAO (111), the presence of additional Bragg reflections (inset in Fig. 6(b)) for the 5 uc film on STO (111) confirmed the appearance of an additional non-perovskite phase during the initial stage of LNO growth. This secondary chemical phase was identified as $La_2Ni_2O_5$ from X-ray diffraction, which was further corroborated by the observation of dominant contribution from $Ni^{2+}$ ions in XA (X-ray absorption) spectrum (Fig. 6(b)). Similar secondary phase formation has been also reported for LNO film on STO (111), grown by molecular beam epitaxy [65]. With the increase of LNO film thickness as the increased metallicity screens the effect of polar jump, the relative amount of $Ni^{3+}$ ions increases for the expense of $Ni^{2+}$ and finally stoichiometric $LaNiO_3$ was obtained at around 15 uc along [111]. On the other hand, $Ni^{3+}$ was stabilized from the very initial stage of growth of LNO on LAO (111) substrate and thus confirmed that the absence of a polar jump at the film/substrate interface was critically important for the epitaxial stabilization along [111]. As a result, the desired generalized graphene like crystal of *RE*$NiO_3$ with *RE* = La to Nd was successfully achieved on LAO (111) substrate [23].

According to the theoretical calculations [12,51], the QAH topological phases should be accompanied by spontaneous ferromagnetism. XRMS (X-ray resonant magnetic scattering) measurements on $2NdNiO_3/4LaAlO_3$ (111) (denoted as 2NNO/4LAO) superlattice, however, ruled out the possibility of a long-range ferromagnetic ground state and instead established the presence of antiferromagnetic correlations. In addition, XAS measurements using linearly polarized X-rays found the presence of large around 9 % XLD (X-ray linear dichroism) on Ni $L_2$ edge, as shown in Fig. 6(c). This XLD signal together with the *ab-intio* calculations implies an unusual antiferro-orbital ordered state, which does not exist in either bulk or (001) oriented nickelate thin films or heterostructures [25]. In short, the artificially stabilized buckled honeycomb lattice structure for various other perovskite based transition metal oxides can be a very promising route for realizing new types of correlated Mott phases or interacting topological phases. In addition, very recently the fabrications of (111)-oriented bilayers and trilayers of transition metal perovskites $SrIrO_3$ have been reported [26,27] with the goal to access new topological phases.

Another interesting application of GLE is connected to designer magnetic phases by layering along [111]. For example, LNO has been heterostructured with $LaMnO_3$ (LMO) along [111] direction [66]. Although bulk LNO is paramagnetic and bulk LMO is antiferromagnetic, the (111)-oriented LNO/LMO superlattices found to exhibit strong exchange bias (EB) effect. We note that EB is generally observed for ferromagnetic/antiferromagnetic or ferromagnetic/spin-glass interface. [67,68] This surprising result was attributed to the quantum confinement effect [69] and charge transfer between Ni and Mn [70], which are more effective for (111) orientation compared to (001). Very recent XAS experiments confirmed the larger electron transfer across the interface for (111) heterostructures. [71].

## V. Artificial double-perovskite lattice

In the past decade ordered double perovskite oxides with the chemical formula $A_2BB'O_6$ (A is an alkali-earth ion, and B and B' are two different transition metal ions) have motivated plethora of research activities. [72] In this family, the B and B' ions are naturally ordered by forming a rock salt sublattice with B-O-B' super-exchange that can generate rich magnetic and topological behaviors depending on the choice of the transition metal elements. Thus the ordered double perovskites offer an excellent engineering playground for which potential functionalities can be predicted and explored [72-76]. From the synthesis perspective, in the bulk the formation of an ordered double perovskite requires a relatively large difference between the chemical valences of B and B' ions. Moreover, the ionic radii difference between those ions must be within a limited range often preventing synthesis of a variety interesting structures. [72] Consequently, there are formidable challenges for the bulk synthesis of many theoretically proposed ordered double perovskite materials which are expected to host many interesting phenomena [20,77-79]

With the recent advances in thin film growth techniques, it was noticed that there is another way to overcome those limitations. [66,80-83] As illustrated in Fig. 7, the structure of an ordered double perovskite $A_2BB'O_6$ consists of alternating monolayers of $ABO_3$ and $AB'O_3$ placed along the (111) orientation. By epitaxially growing (111)-oriented $1ABO_3/1AB'O_3$ superlattices, it is now possible to design a wide range of ordered double perovskite compounds that have never been realized through traditional bulk synthesis. In contrast to the artificial graphene-like lattices, here the out-of-plane stacking sequence of each perovskite composition has to be maintained with a monolayer precision to realize the required double perovskite structure.

Following this idea, a series of artificial double perovskite $La_2BB'O_6$ (B and B' = Fe, Cr, Mn, and Ni) in the thin film form were obtained by PLD [66,81,82]. Among these materials, the artificial $La_2FeCrO_6$ ($1LaFeO_3$-$1LaCrO_3$) system has been extensively investigated and yet some controversy about its electronic and magnetic structure remained due to complexity of $Fe^{3+}$-O-$Cr^{3+}$ superexchange. [84]. Ueda *et al.* [81] grew the artificial $La_2FeCrO_6$ (LFCO) films for the first time and found the samples to show ferromagnetic behavior. However, the small magnetic moment of only ~ 0.009 $\mu_B$ at 0.1 T raised questions about the true nature of the magnetic ordering; to complicate the case both canted antiferromagnetic and ferrimagnetic ground states have been theoretically predicted. [85]

To resolve the controversy, Ben *et al.* [86] synthesized artificial $La_2FeCrO_6$ and performed XAS and XMCD measurements at the *L*-edge of both Fe and Cr ions to probe the electronic and magnetic properties of each transition metal element. The corresponding absorption data shown in Fig. 8 confirmed that in the artificial LFCO compound both Fe and Cr are trivalent with high-spin configurations ($Fe^{3+}$: S = 5/2; $Cr^{3+}$: S = 3/2). In addition, the XMCD spectra of both elements revealed that the magnetic moments of $Fe^{3+}$ and $Cr^{3+}$ ions are parallel to each other. The estimated magnetic moments were about 0.3 $\mu_B$/Cr and 0.2 $\mu_B$/Fe at 5 T and appeared to be much smaller than the expected saturated values (3 $\mu_B$/Cr and 5 $\mu_B$/Fe). Furthermore, no detectable XMCD signal was observed for either ion under 0.1 T. Based on those observations, the artificial $La_2FeCrO_6$ was claimed to have a canted AFM ground state. Chakraverty *et al.* [87] studied the relation between the degree of B site disorder and the corresponding magnetic state on a series of thick LFCO films (up to a few hundred nanometers) that were synthesized with a LFCO solid-solution target. They observed that only the high quality B-site ordered sample show the expected saturated magnetization of ~ 2 $\mu_B$ per LFCO formula, consistent with a ferrimagnetic ordering picture in which the local spin moment of Cr and Fe ions are anti-aligned, while the samples with sizable B, B'-site cation disorder show strong suppression of the total magnetic moment.

In addition, $1BiFeO_3/1BiCrO_3$ (BFCO) artificial double perovskite superlattice was also developed by Ichikawa *et al.* [80] on STO (111) substrate and room temperature multiferroic behavior was observed on this system. Interestingly, the saturated magnetic moment per formula unit of the superlattice was much stronger than the signal of the film grown from the BFCO sputtering target [88]. This enhanced magnetization due to GLE sheds additional light on its potential for the development of new magnetic materials for application.

## VI. Summary and outlook

In the previous sections, the current status of GLE is discussed mainly with emphasis on the system layered along the cubic (or pseudo-cubic) (111) orientation of perovskites and spinels. However, the GLE concept is certainly not restricted to those systems and its realization opens wide opportunities for experimentalists to fabricate other interesting lattices in the heterostructured form. For instance, apart from spinel-type compounds, 2D frustrated lattices also appear in ultra-thin (111)-oriented pyrochlore (chemical formula: $A_2B_2O_7$) structures (Fig. 9, right panel on the top). This can be easily understood since the B sublattice in the pyrochlore structure is identical to that of spinel. Several theoretical works [12,17,18,89] have predicted the emergence of topological states with different B cations, while the experimental work on fabricating (111) pyrochlore films have been recently initiated [90-93]. Additionally, it was theoretically proposed that a monolayer of corundum-type (chemical formula: $M_2O_3$) structure grown along the (0001) direction could also form the graphene-like honeycomb lattice geometry [94] analogous to the perovskite (111) bilayers (Fig. 9, left panel on the top).

To push this field further, one can envision that a combination of GLE with other engineering approaches should allow for even more exotic phases. A fusion of GLE and interface engineering may enable new tailor-made phenomena since for heterostructures the mismatches at the interface should be profoundly altered by the epitaxial orientation. For instance, while the basic scenario of two-dimensional electron gas (2DEG) at the interface between two insulating complex oxides has been interpreted within the 'polar discontinuity' model [95,96], the polar mismatch patterns vary with the orientation of the heterostructure. This can be utilized to tune the carrier density and mobility of the 2DEG.

Towards experimental realization of this idea a pioneering work has been reported on the growth of the prototypical $LaAlO_3/SrTiO_3$ system with (001), (110), and (111) interfaces (Fig. 9, right panel on the bottom) [97]. Another interesting theory proposal is the study on the interface between (100)-oriented high-temperature superconducting cuprate $YBa_2Cu_3O_7$ (YBCO) and colossal magnetoresistance (CMR) manganite $La_xCa_{1-x}MnO_3$ (LCMO). [98] Compared to the extensively investigated (001) YBCO/LCMO interface [99] where the nodes of Cu $d_{x2-y2}$ orbital are in the a-b plane, the (100) oriented YBCO has nodes of the $d_{x2-y2}$ orbital out of the $CuO_2$ plane (see Fig. 9, left panel on the bottom). This unique orbital coupling pattern is crucial for the emergence of exotic Majorana states [100]. Furthermore, the combination of GLE and strain engineering may open another interesting dimension for controlled modifications of the bond-length and bond-angle in structural units by epitaxial strain along the unconventional crystallographic axes. [100-102]

To summarize, with the modern state-of-the-art thin film fabrication methods, the notion of GLE can be experimentally realized and used as another control knob. Despite the challenges of layering with atomic precision along unconventional crystallographic directions, GLE has excellent potential for the discovery of novel electronic, magnetic and topological phases with complex oxides.

**Acknowledgments**

The authors would like to thank D. Khomskii and G. Fiete for enlightening discussions. J.C. was supported by the Gordon and Betty Moore Foundation's EPiQS Initiative through Grant No. GBMF4534. X.L. and S.M. were supported by the Department of Energy under Grant No. DE-SC0012375, and Y.C. and M.K. were supported by the DOD-ARO under Grant No. 0402-17291.

Figure 1 Schematic illustration of the idea about geometrical lattice engineering, which has three control parameters named as orientation, stacking, and superlattice throughout its entire process.

Figure 2 The conventional unit cell of a normal spinel structure. The corresponding (111) ionic planes are marked on the figures. Note, the A ionic planes only form the triangle planes while the B ionic planes form both the triangle planes but also the kagome planes.

Figure 3 (111)-oriented $CoCr_2O_4/Al_2O_3$ (CCO/AlO) heterostructures. (a) Epitaxial relationship between CCO and AlO. The hexagonal close packing of AlO oxygen sublattice is labeled as AB in black, while the cubic close packing of CCO oxygen sublattice is labeled as ABC in brown. (b)-(d) RHEED images during the growth of each component. The half order reflections (marked by pink solid circles) observed on CCO layer is due to the double expansion of the in-plane unit cell. Reproduced from Ref. [24], with the permission of AIP Publishing.

Figure 4 (a) X-ray reflectivity and (b) X-ray diffraction curvevs of 2CCO/nAlO (n = 1 and 2) superlattices. Both the thickness fringes and the superlattice satellite peaks are clearly seen from the graphs. (c) and (d) X-ray absorption spectra of the CCO/AlO superlattices on the $L_{2,3}$ edges of Co and Cr, respectively. The spectra of CCO powders with normal spinel structure are plotted as references (Figure 4 (a) and 4 (b) adapted with permission of AIP Publishing from Ref. [46]).

Figure 5 Buckled honeycomb lattice and Dice lattice geometry can be generated by epitaxial growth of perovskite system along [111] direction. *A* sites are not shown for Dice lattice due to visual clarity.

Figure 6 (a) Schematics (assuming pure ionic picture) to demonstrate the presence (absence) of polar jump for *RE*NiO3 on $SrTiO_3$ ($LaAlO_3$) (111) substrate [61]. (b) XAS of 5 uc LNO films on LAO and STO (111) have been compared with bulk $Ni^{2+}O$, and $LaNi^{3+}O_3$. RHEED pattern of these films, recorded along pseudocubic [1 -1 0]. For details, see Ref. [63]. (c) Experimental arrangement for measuring XLD of (111) oriented superlattice (upper panel) and XA spectra recorded for a [$2NdNiO_3/4LaAlO_3$]x3 SL grown on LAO (111) using vertically (V) and horizontally (H) polarized X-ray and their difference are plotted in lower panel. Due to strong overlap of Ni $L_3$ edge with La $M_4$ edge (from the substrate and spacer layer LAO), only Ni $L_2$ edge is shown. Figure 6 (c) adapted with permission from Ref. [25]. Copyrighted by the American Physics Society.

Figure 7 B-site ordered double perovskites viewed along the conventional (001) and the unconventional (111) direction, respectively. The (001) rock-salt arrangement of the B and B' sites is equivalent to (111) $BO_6$ and $B'O_6$ superlattice.

Figure 8 XAS and XMCD spectra on the L edge of (a) Fe and (b) Cr of the (111)-oriented $1LaFeO_3/1LaCrO_3$ superlattice. Both spectra were measured at 10 K with an external magnetic field ~ 5 T applied parallel to the film surface. Data were collected in the total electron yield mode. Adapted from Ref. [86], with the permission of AIP Publishing.

Figure 9 Summary and outlook on GLE. The up panel displays other possible routines to topological phase and frustrated magnetism by applying pure GLE. The bottom panel presents the combination of GLE with IE or SE to establish new systems with intriguing physics such as quasi-particle excitation and emergent phenomena.

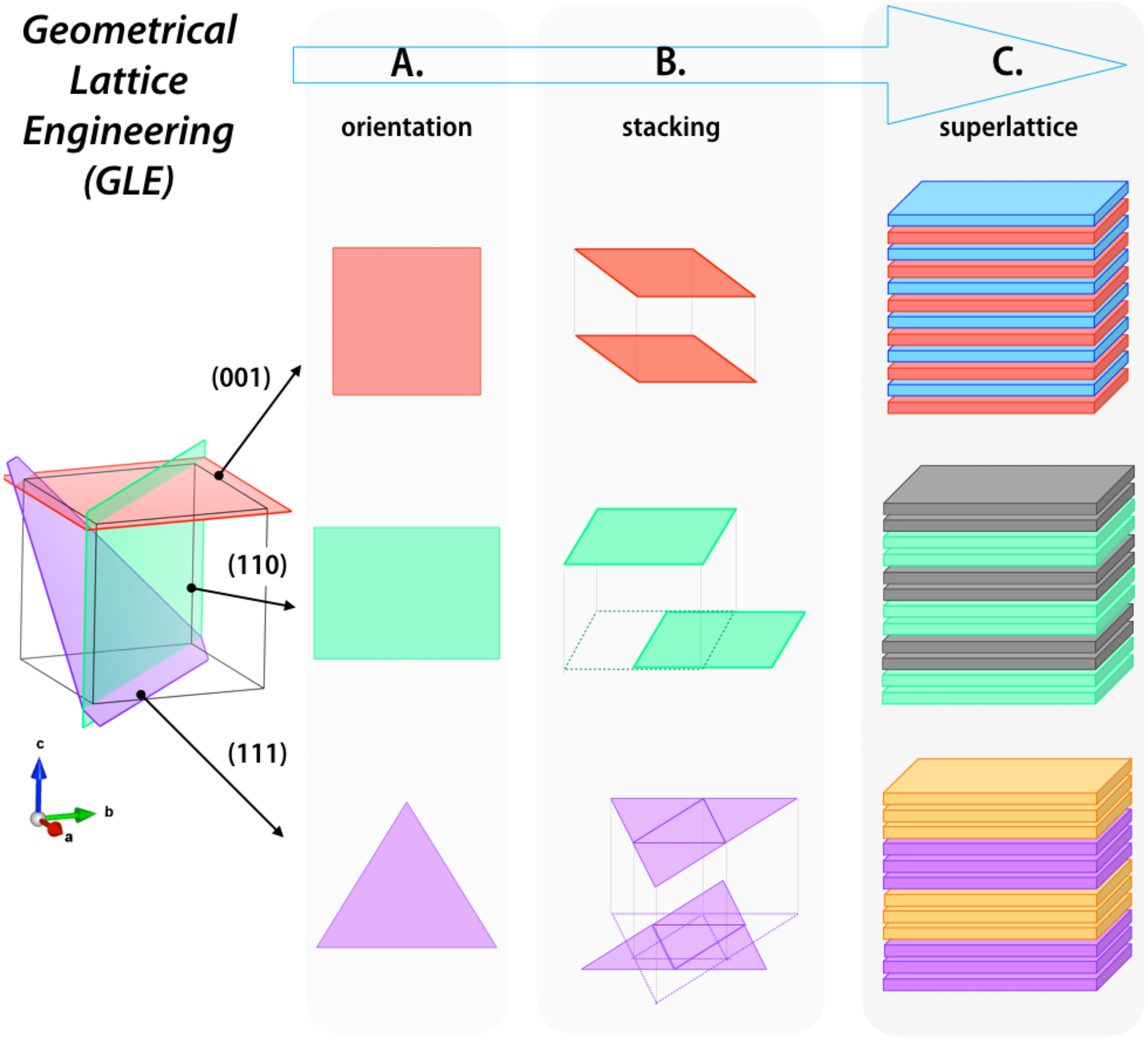

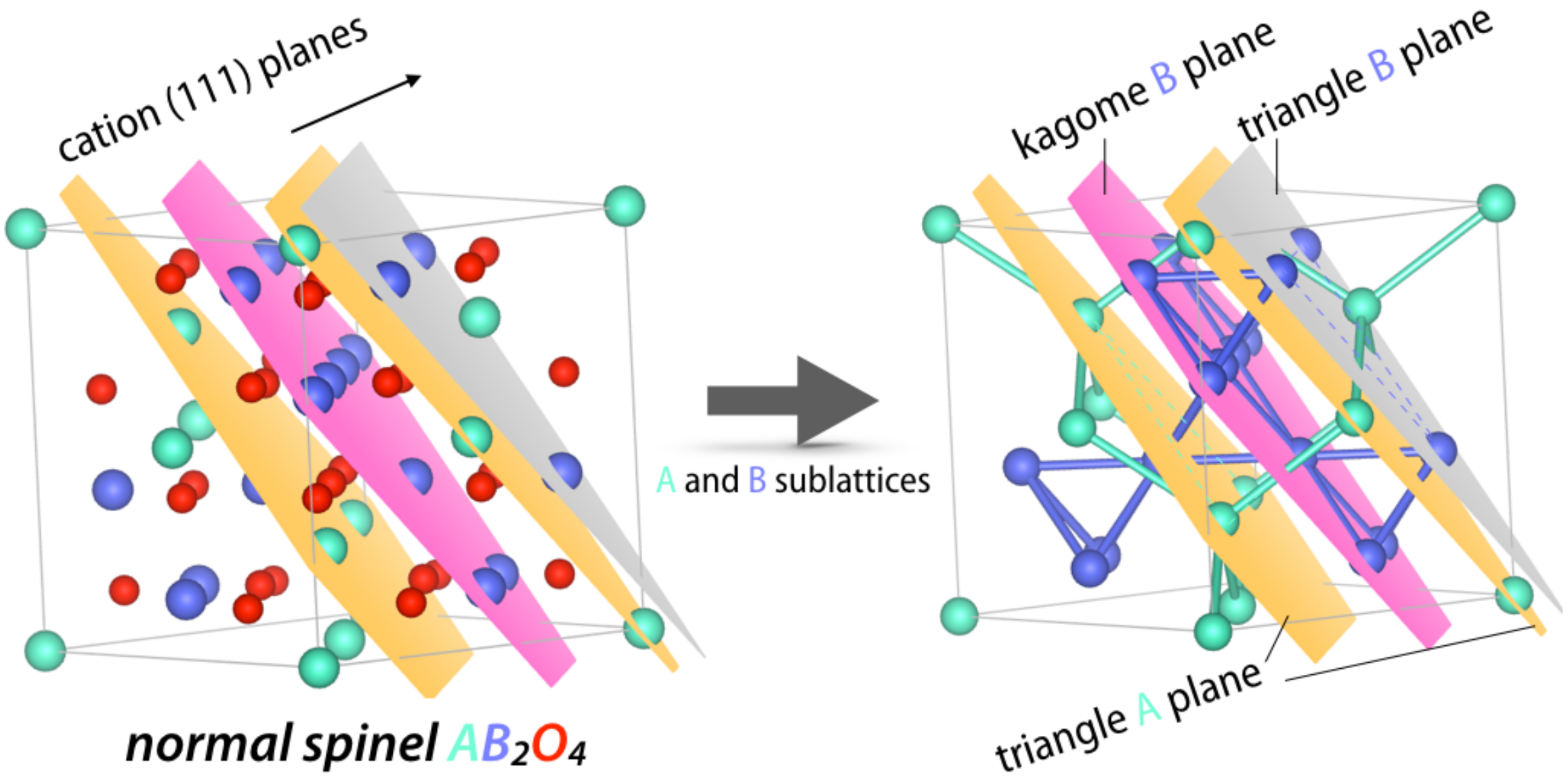

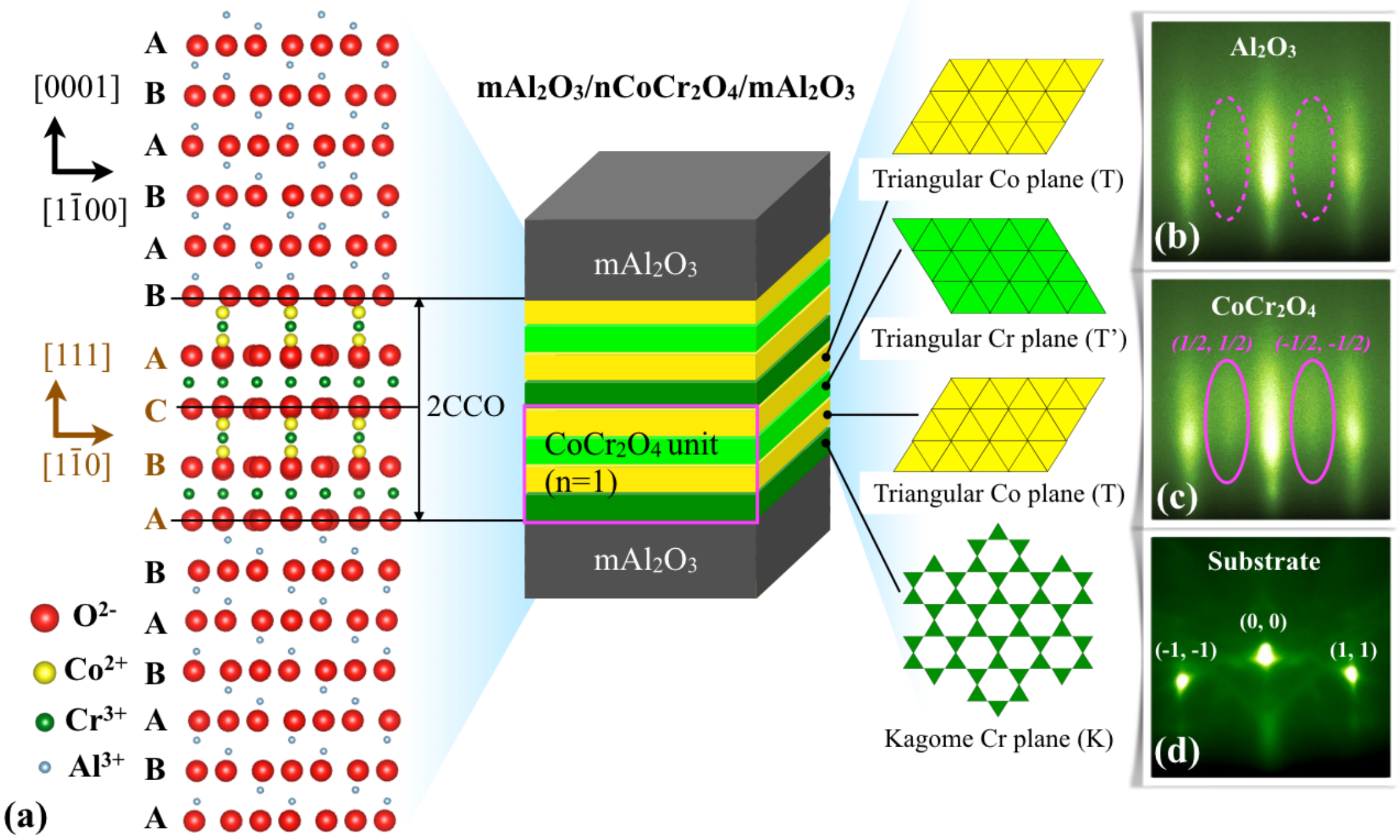

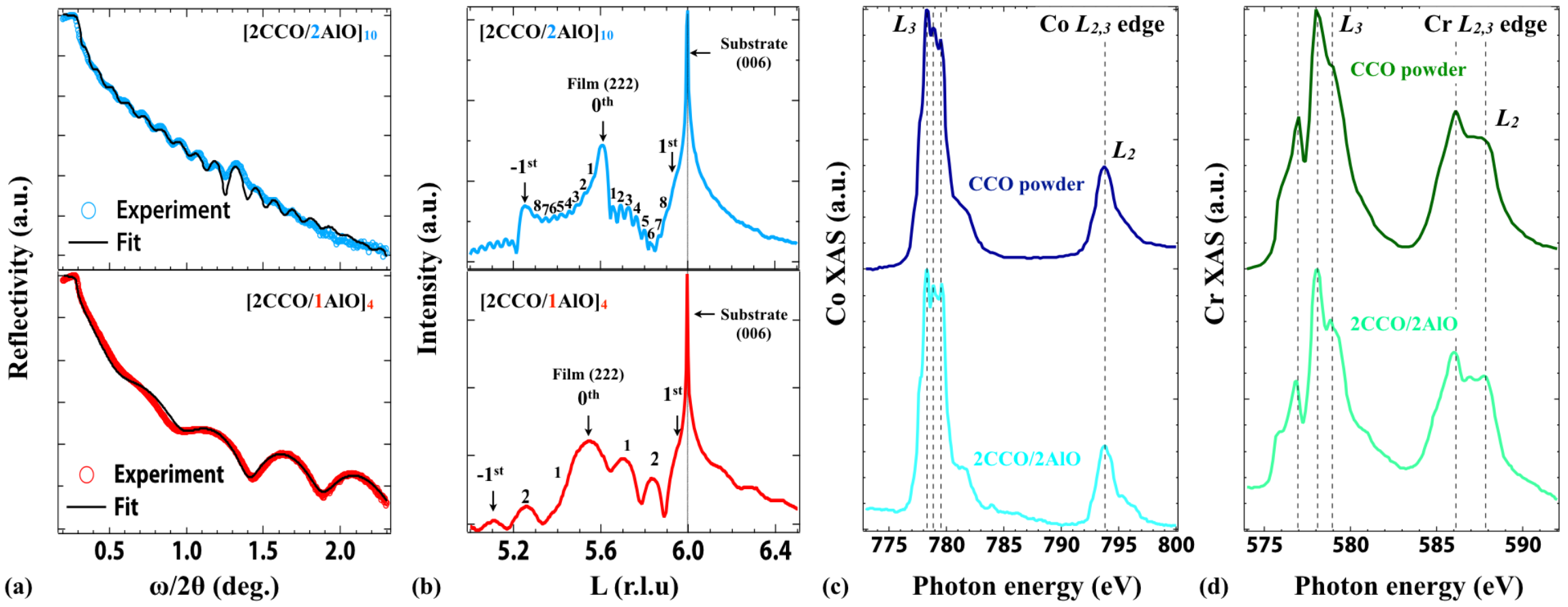

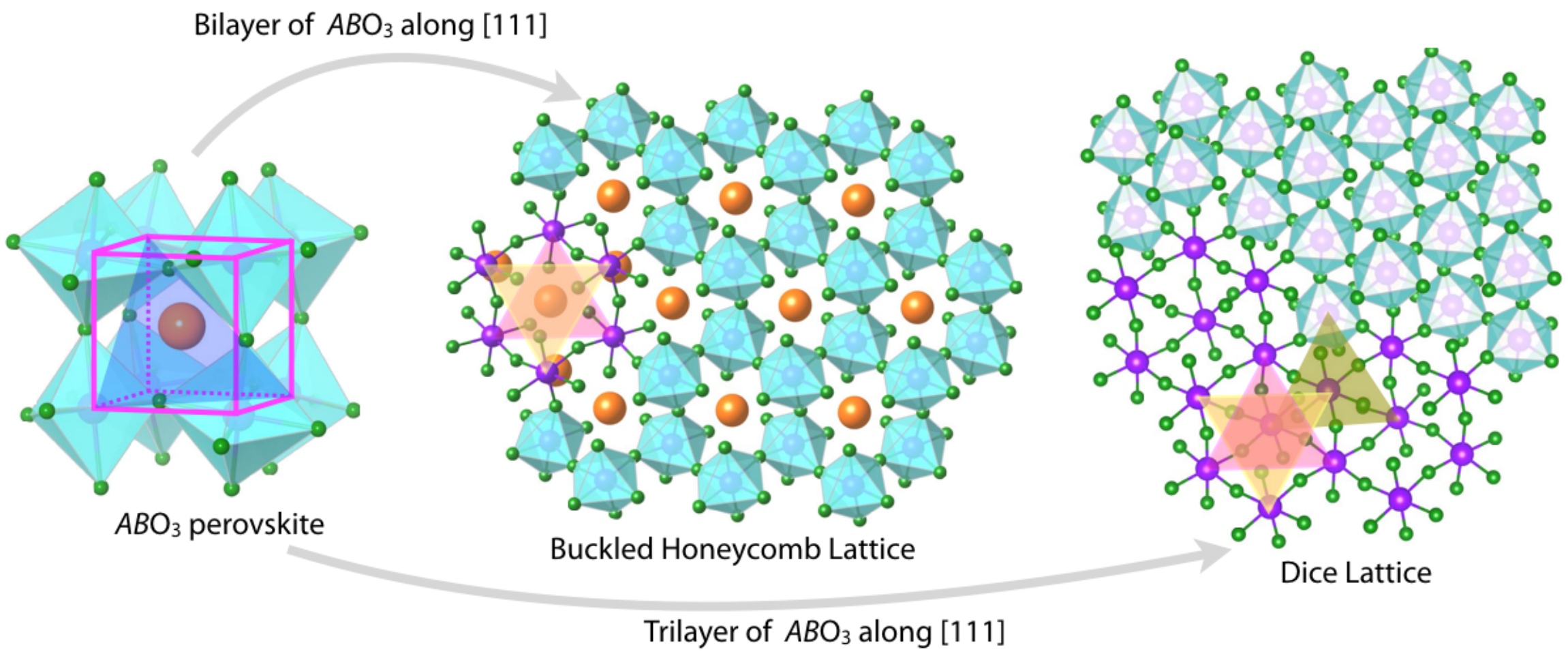
Bilayer of ABO3 along [111]
ABO3 perovskite
Buckled Honeycomb Lattice
Dice Lattice
Trilayer of ABO3 along [111]

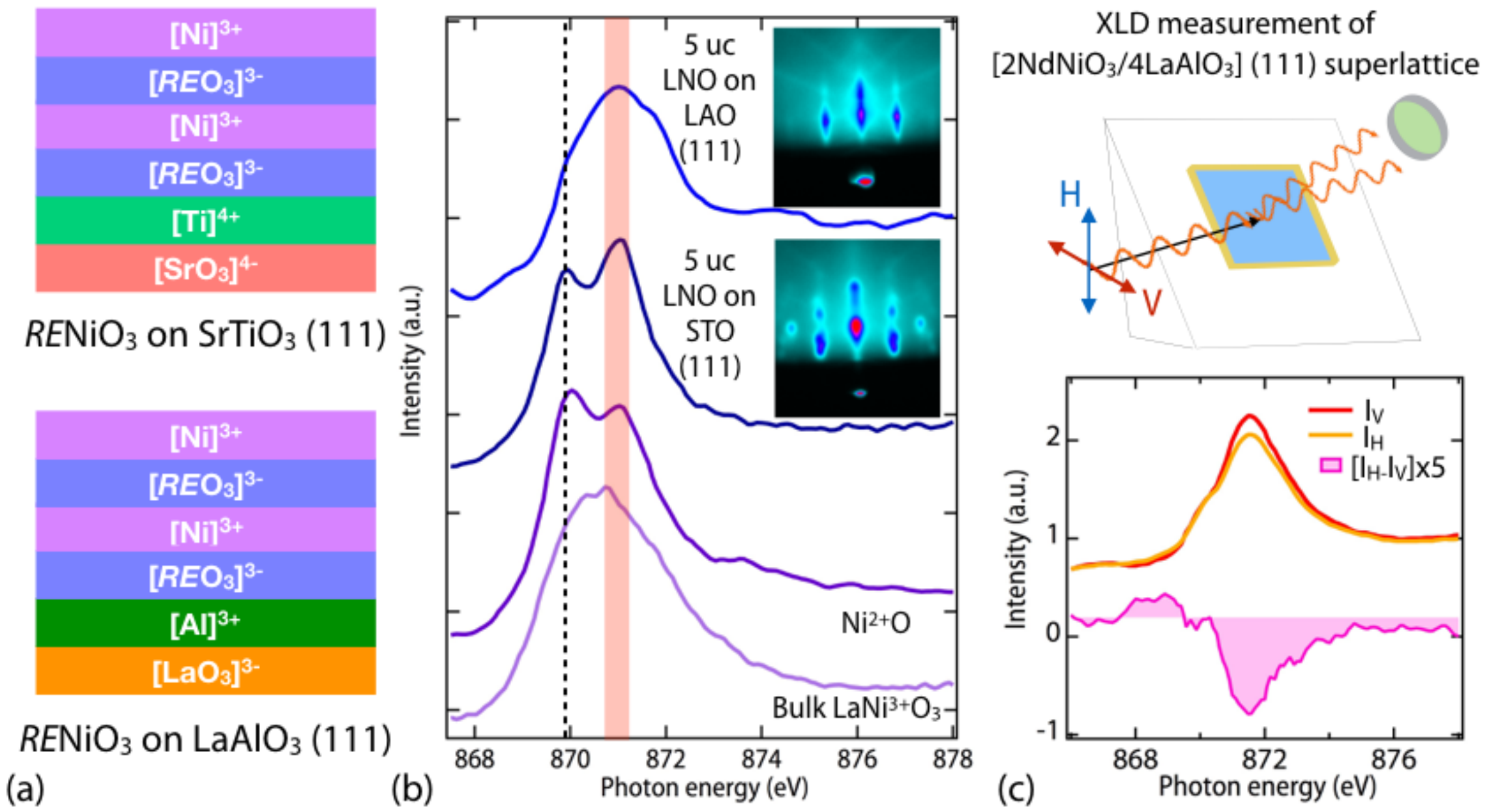

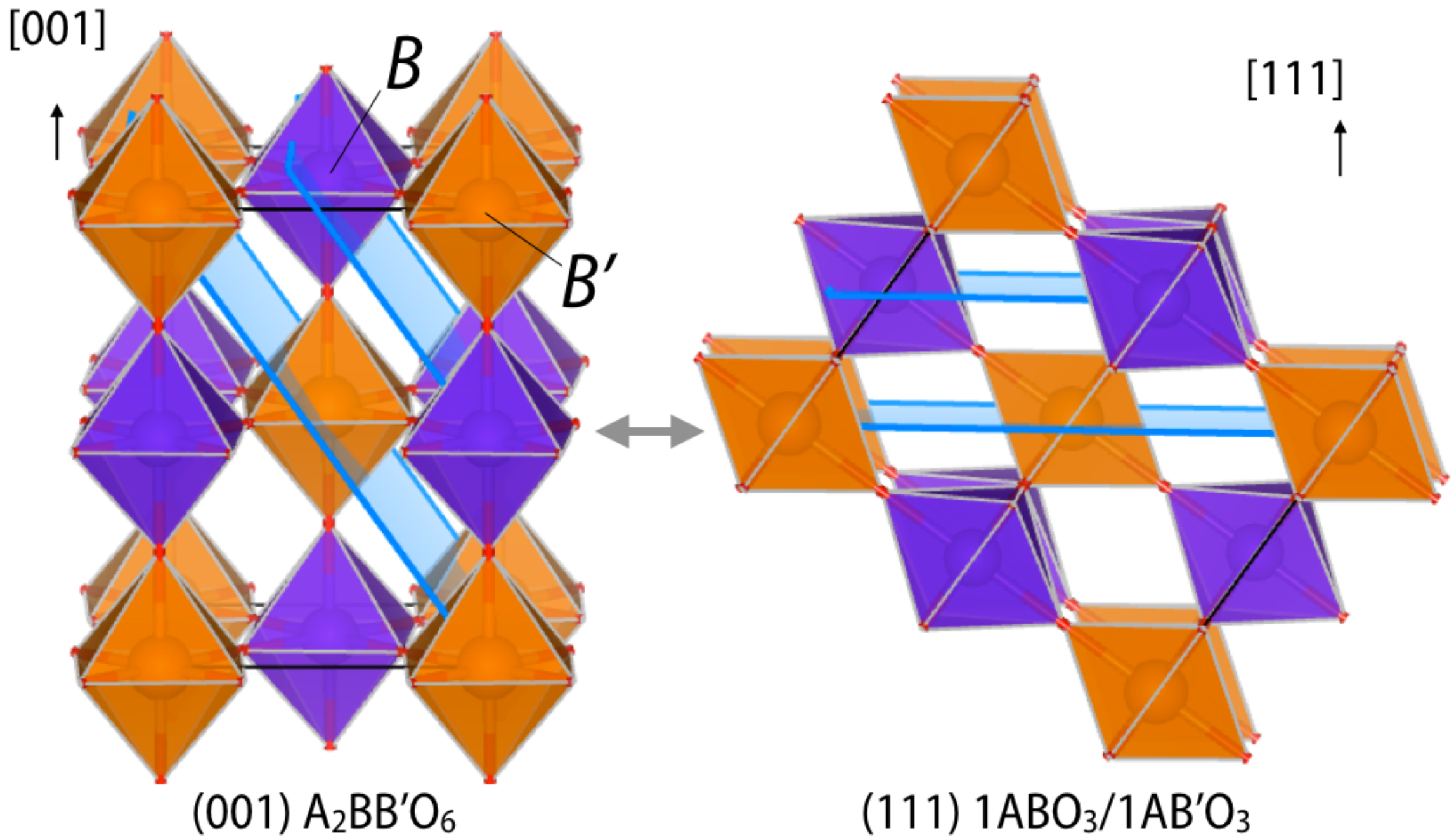

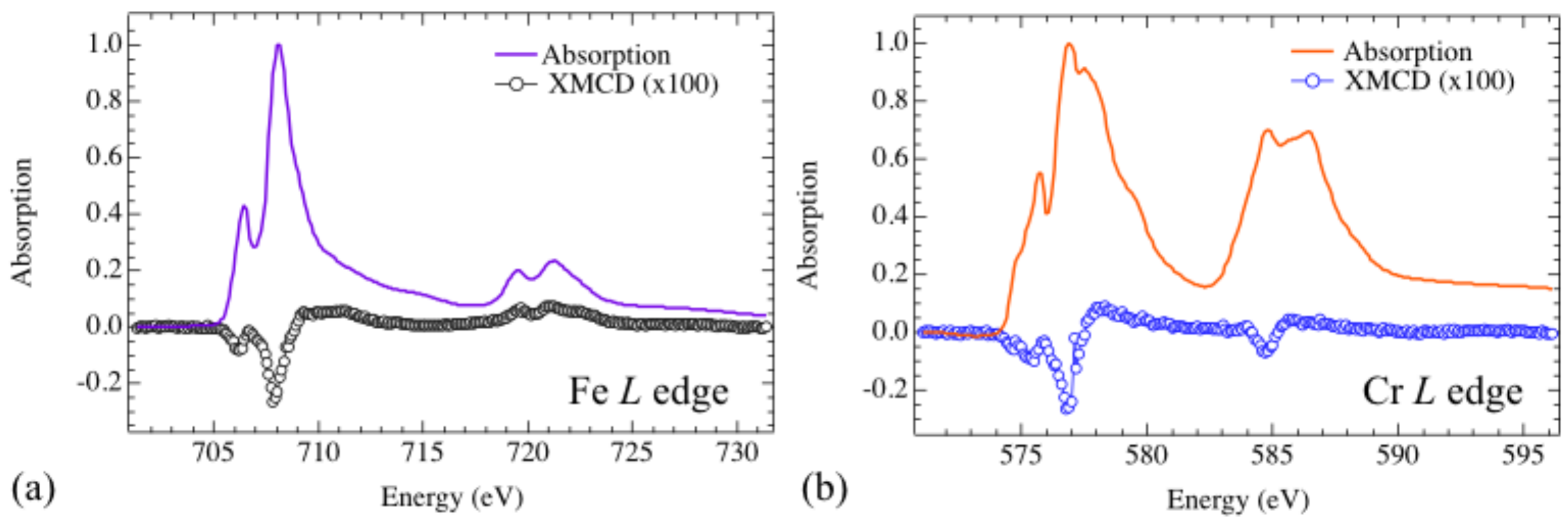

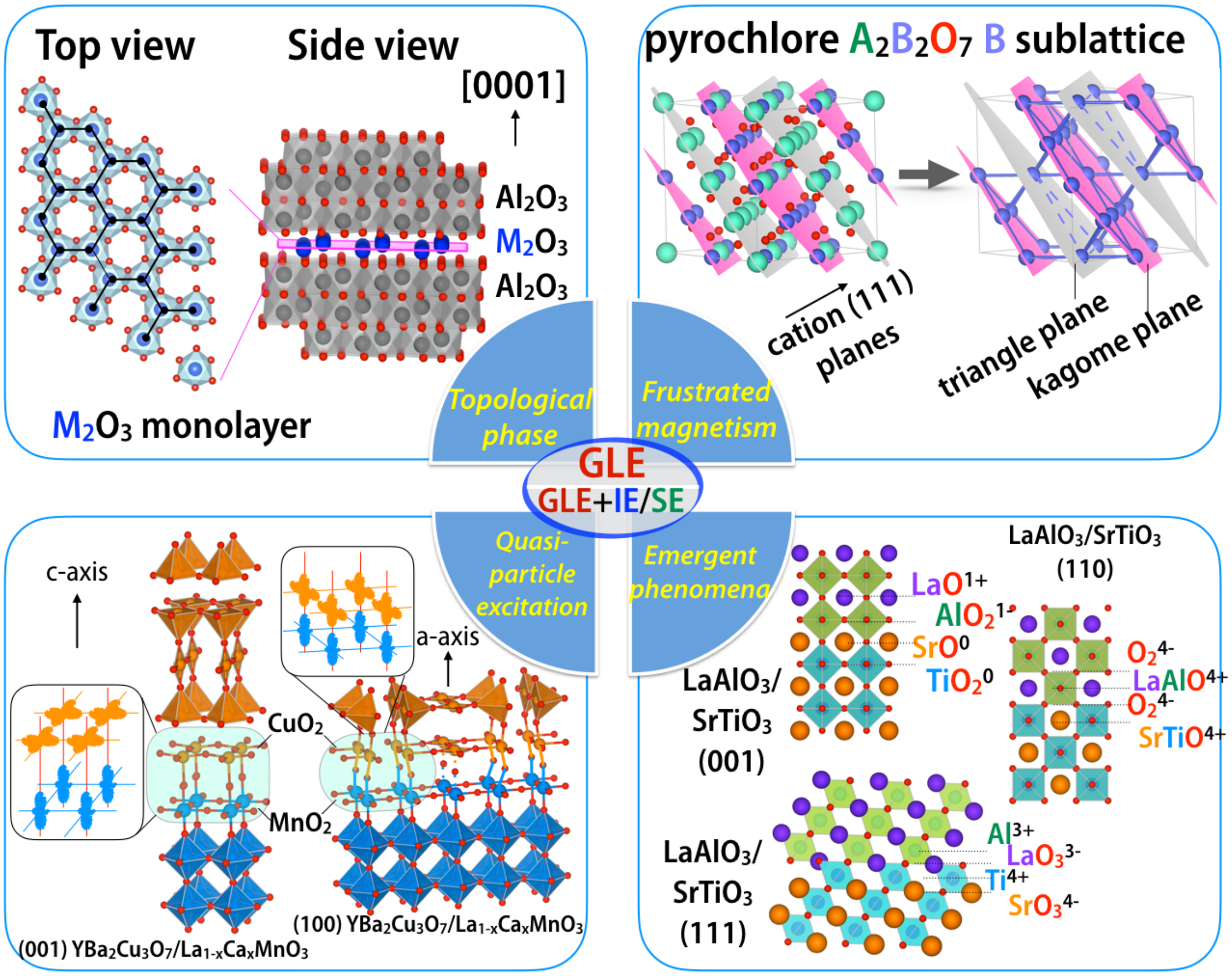